\documentstyle[multicol,pre,aps]{revtex}
\input psfig
\draft

\begin{document}

\title{Thermodynamic and structural aspects of the \\potential energy
surface of simulated water}

\author{Francis~W. Starr$^{1,2}$, Srikanth Sastry$^3$, Emilia La
Nave$^{1}$, Antonio Scala$^{1,4}$, \\ H. Eugene Stanley$^1$, and Francesco
Sciortino$^4$}

\address{$^1$Center for Polymer Studies, Center for Computational
Science, and Department of Physics, Boston University, Boston, MA
02215 USA}

\address{$^2$ Polymers Division and Center for Theoretical and
Computational Materials Science, National Institute of Standards and
Technology, Gaithersburg, MD, 20899 USA}

\address{$^3$ Jawaharlal Nehru Centre for Advanced Scientific Research, 
Jakkur Campus, Bangalore 560064, INDIA}

\address{$^4$Dipartmento di Fisica e Istituto Nazionale per la Fisica
della Materia, Universit\'{a} di Roma ``La Sapienza'', Piazzale Aldo
Moro 2, I-00185, Roma, ITALY}

\date{28 July 2000}

\maketitle

\begin{abstract}

Relations between the thermodynamics and dynamics of supercooled
liquids approaching a glass transition have been proposed over many
years. The potential energy surface of model liquids has been
increasingly studied since it provides a connection between the
configurational component of the partition function on one hand, and
the system dynamics on the other.  This connection is most obvious at
low temperatures, where the motion of the system can be partitioned
into vibrations within a basin of attraction and infrequent
inter-basin transitions.  In this work, we present a description of
the potential energy surface properties of supercooled liquid water.
The dynamics of this model has been studied in great details in
the last years.  Specifically, we locate the minima sampled
by the liquid by ``quenches'' from equilibrium configurations
generated via molecular dynamics simulations.  We calculate the
temperature and density dependence of the basin energy, degeneracy,
and shape.  The temperature dependence of the energy of the minima is
qualitatively similar to simple liquids, but has anomalous density
dependence.  The unusual density dependence is also reflected in the
configurational entropy, the thermodynamic measure of degeneracy.
Finally, we study the structure of simulated water at the
minima, which provides insight on the progressive tetrahedral ordering
of the liquid on cooling.

\end{abstract}


\begin{multicols}{2}

\section{Introduction}

In recent years, numerical study of model liquids in supercooled states
has been helpful to clarify the physics of the glass
transition~\cite{kob-review}.  The availability of long trajectories in
phase space offers the possibility of closely examining the changes in
supercooled states that are responsible for slowing down the dynamics by
15 decades in a narrow temperature range approaching the glass
transition.  Although current computational studies are limited to times
shorter than $\approx 1$~$\mu$s (as opposed to real liquids, for which
the dynamics can be studied up to $\approx 10^2$~s), a coherent picture
of the glass transition phenomenon is beginning to emerge.

In addition to characterizing changes in the dynamics, recent studies
have demonstrated the utility of examining the underlying potential
energy surface (PES) as an aid to understanding the properties of
supercooled liquids connecting the dynamics to the thermodynamics and
the topology of configuration
space~\cite{sri-nature,ruocco,skt,heuer,coluzzi,sslss-nature,tbs,sri-prl}.
This connection is most obvious at low temperatures, where the motion of
the system can be partitioned into motion confined within a single
potential energy basin with infrequent inter-basin transitions.  It has
been shown that at sufficiently high temperature (at constant volume),
the system explores always the same distribution of basins, and that the
average basin energy is nearly temperature independent. Below a
crossover temperature---which is coincident with the onset of a two-step
relaxation in the decay of density fluctuations \cite{sri-nature}---the
system starts to populate basins of progressively lower energy, but
which are less numerous.

The thermodynamic approach based on the analysis of the PES, following
the formalism proposed by Stillinger and Weber\cite{stillinger}, has
become a powerful formalism for the interpretation of numerical data,
both in
equilibrium~\cite{sri-nature,ruocco,skt,heuer,coluzzi,sslss-nature,tbs,sri-prl}
and in out-of-equilibrium conditions~\cite{eplaging,st-aging}.  The
degeneracy of the energy minima, i.e., the number of basins with a
selected minimum energy, has been quantified for several model systems
and used to calculate the configurational entropy $S_{\mbox{\scriptsize
conf}}$, from which an ``ideal'' glass transition (in the sense of
Kauzmann, Adam, Gibbs, and DiMarzio~\cite{kauzmann,gibbs,ag}) has been
estimated~\cite{skt,heuer,sslss-nature,sri-prl}.

In this paper we present a detailed investigation of the properties of
local potential energy minima, or ``inherent structures'' (IS), sampled
by the extended simple point change (SPC/E) model of water~\cite{spce}.
Previous studies on the PES for models of water have shown the relevance
of this approach to the deepening of our understanding of both
structural and dynamical properties of liquid
water~\cite{sslss-nature,still,ohmine-tanaka,fsquench,sciortino2,ohmine,rds99,sasai}.
Specifically, we calculate the temperature and density dependence of the
basin energy over a wide range of temperatures and densities.  We also
study in detail the shape of the basins in configuration space and
estimate their degeneracy.  The information presented provides a
detailed characterization of the PES and furnishes all information
required to explicitly write the liquid Helmholtz free energy in a wide
temperature and density range.  Finally, we study the geometrical
arrangement of the molecules at IS minima to better understand the
changes that take place in the liquid on cooling.  Focusing on the IS
allows us to eliminate thermal effects which complicate the temperature
dependence.  In particular, we focus on the fraction of water molecules
that are four-coordinated.

\section{Simulations}

The majority of the state points studied here are from the molecular
dynamics simulations of the SPC/E model performed in
Ref.~\cite{shss}. The simulation methods are discussed in
Ref.~\cite{shss}.  We have performed additional simulations of ice
$I_h$, so that we can compare the IS properties of the crystal with
those of the liquid.  The simulations of ice consist of a periodic box
containing 432 molecules with dimensions $2.634~\mbox{nm} \times
2.281~\mbox{nm} \times 2.151$~nm for density $\rho = 1.0$~g/cm$^3$.
The dimensions are uniformly scaled in order to obtain other
densities.  The box dimensions have been optimized to generate the
lowest energy configuration at density $1.0$~g/cm$^3$.  Proton
disorder in the initial configuration is generated by identifying
closed hydrogen bond loops, and exchanging hydrogens between
molecules, as described in Ref.~\cite{stillinger-ice}.

The ice simulations have been performed at $\rho = 0.90$, 0.95, 1.00,
and 1.05~g/cm$^3$ and temperature $T = 194$~K.  The thermodynamic
properties are summarized in Table~\ref{table:ice-simulations}.  The
equilibration time for these sample is far less than that of the liquid
at the same temperature since only the vibrational degrees of freedom
need to be relaxed.

We have performed conjugate gradient minimizations
\cite{numerical-recipes} to locate local minima on the PES closest to
any given instantaneous configuration.  We use a tolerance of
$10^{-15}$~kJ/mol in the total energy for the minimization.  For each
state point, we quench at least 100 configurations taken from two
independent trajectories.  While each configuration is not necessarily
separated by the typical relaxation time of the system, the set of
points quenched typically spans $\approx 25$ times the relaxation time
of the intermediate scattering function.  

\section{Inherent Structure Properties of SPC/E}

\subsection{Thermodynamics in the Inherent Structure Formalism}
\label{sec:II}

Stillinger and Weber formalized the concept of a basin in the potential
energy surface by introducing the inherent structure formalism
\cite{stillinger}. The set of points that map to the same minimum via
steepest descent are those which constitute a basin, and the minimum of
a basin is the IS.  This approach is particularly well suited to
simulated liquids, since it is possible to explicitly calculate the
steepest descent trajectory from an equilibrium state point.  Moreover,
the partition function $Z$ can be explicitly written in terms of the
basins.  In the isochoric-isothermal (NVT) ensemble, for a system of $N$
rigid molecules

\begin{equation}
Z = \lambda^{-6N} \int \exp (- V/k_BT)d^N{\bf r}
\end{equation}

\noindent which can be written as a sum over all basins in
configurational space, i.e.
\begin{eqnarray}
Z &=& \lambda^{-6N} {\sum_{\mbox{\scriptsize basins}}} \exp
(-e_{IS}/k_BT) \times \\
& & ~~~~~\int_{R_{\mbox{\scriptsize basin}}} \exp (-(V-e_{IS})/k_BT) d^N{\bf
r}. \nonumber
\label{eq:partition-fctn}
\end{eqnarray}
Here $\lambda\equiv h (2\pi mk_B T)^{-1/2}$ is the de Broglie
wavelength, $V \equiv V({\bf r}^N)$ is the potential energy as a
function of the atomic coordinates, $e_{IS}$ is the energy of the IS and
$R_{basin}$ is the configuration space associated to a specific basin.
The model system we consider here, namely SPC/E water, has six degrees
of freedom for each molecule.  It is natural to introduce
$\Omega(e_{IS})$, the number of minima with energy $e_{IS}$, and the
free energy of a basin with basin energy $e_{IS}$ $f(T, e_{is})$ (the
``basin free energy'').
\begin{eqnarray}
\label{eq:basin-energy}
f(T,e_{IS}) & \equiv & -k_BT\ln \left( \frac{1}{\Omega(e_{IS})}
\lambda^{-6N} \times \right. \\
 & & \left. {\sum_{\mbox{\scriptsize basins}}}^*
\int_{R_{\mbox{\scriptsize basin}}} \exp [-(V-e_{IS})/k_BT]
d^N{\bf r}\right), \nonumber
\end{eqnarray}
\noindent The asterisk denotes the fact that the sum is constrained to
basins of energy $e_{IS}$. Eq.~(\ref{eq:basin-energy}) accounts for both
the basin structure surrounding the minimum and the kinetic degrees of
freedom.  The complete partition function can be written as the sum over
all possible $e_{IS}$ values,
\begin{equation}
Z = \sum_{e_{IS}} \Omega(e_{IS})
\exp\left({-\frac{e_{IS}+f(T,e_{IS})}{k_BT}}\right)
\end{equation}
or
\begin{equation}
Z= \sum_{e_{IS}} \exp\left({-\frac{-T S_{\mbox{\scriptsize conf}}(e_{IS})+
e_{IS}+f(T,e_{IS})}{k_BT}}\right)
\label{eq:zpoefe}
\end{equation}
where the configurational entropy $ S_{\mbox{\scriptsize
conf}}(e_{IS})\equiv k_B \ln (\Omega (e_{IS}))$.  In the thermodynamic
limit, the corresponding Helmholtz free energy $F(T,V)$ is given by
\begin{equation}
\label{e6}
F(V,T)=  E_{IS}(T) + f(T,E_{IS}(T))- T S_{\mbox{\scriptsize conf}}(E_{IS}(T)) 
\end{equation}
where $E_{IS}(T)$ is the thermodynamic average of $e_{IS}$ and solves
\begin{eqnarray}
\label{eq:freeene}
\partial F(V,T)/\partial e_{IS} & = & 1 + \partial f(T,e_{IS}) /\partial
e_{IS} - \\ & & T \partial S_{\mbox{\scriptsize conf}}(T,e_{IS})/\partial
e_{IS} =0. \nonumber
\end{eqnarray}
$E_{IS}(T)$ can be numerically calculated by estimating the $IS$ which
are populated by a system in equilibrium at temperature $T$ and fixed
volume $V$. Hence, if a good model for $f(T,e_{IS})$ is available,
then the $e_{IS}$ dependence of $S_{\mbox{\scriptsize conf}}$
along isochores can be estimated by integrating Eq.~(\ref{eq:freeene}).
Note that Eq.~(\ref{eq:freeene}) shows that, if the basin free energy
does not depend on $e_{IS}$, then the configurational entropy is
the only quantity controlling the $T$-dependence of $E_{IS}$. In other
words, the statistical mechanics of the basins completely decouples from
the vibrational dynamics~\cite{fstrieste,robin}.

$S_{\mbox{\scriptsize conf}}(E_{IS})$ can also be calculated by studying
the probability distribution $P(E_{IS},T)$, i.e. the probability that
the liquid---in equilibrium at temperature $T$---populates the inherent
structure $E_{IS}$.  Indeed, from Eq.~(\ref{eq:zpoefe})
\begin{eqnarray} 
\label{eq:P(E_IS)}
S_{\mbox {\scriptsize conf}}(e_{IS}) &=& k_B \ln P(e_{IS},T) + e_{IS}/T +
\\ & & f(e_{IS},T) + k_B \ln Z(T). \nonumber
\end{eqnarray}
Hence, Eq.~\ref{eq:P(E_IS)} gives $S_{\mbox {\scriptsize conf}}$ up to an unknown constant
$k_B \ln Z(T)$. This ``histogram technique'' has been recently used to
estimate the configurational entropy for a binary-mixture Lennard-Jones
system~\cite{skt,coluzzi,sri-prl,sritrieste}.

\subsection{Inherent Structure Energy as a function of $\rho$ and $T$}

\begin{figure}[htbp]
\begin{center}
\mbox{\psfig{figure=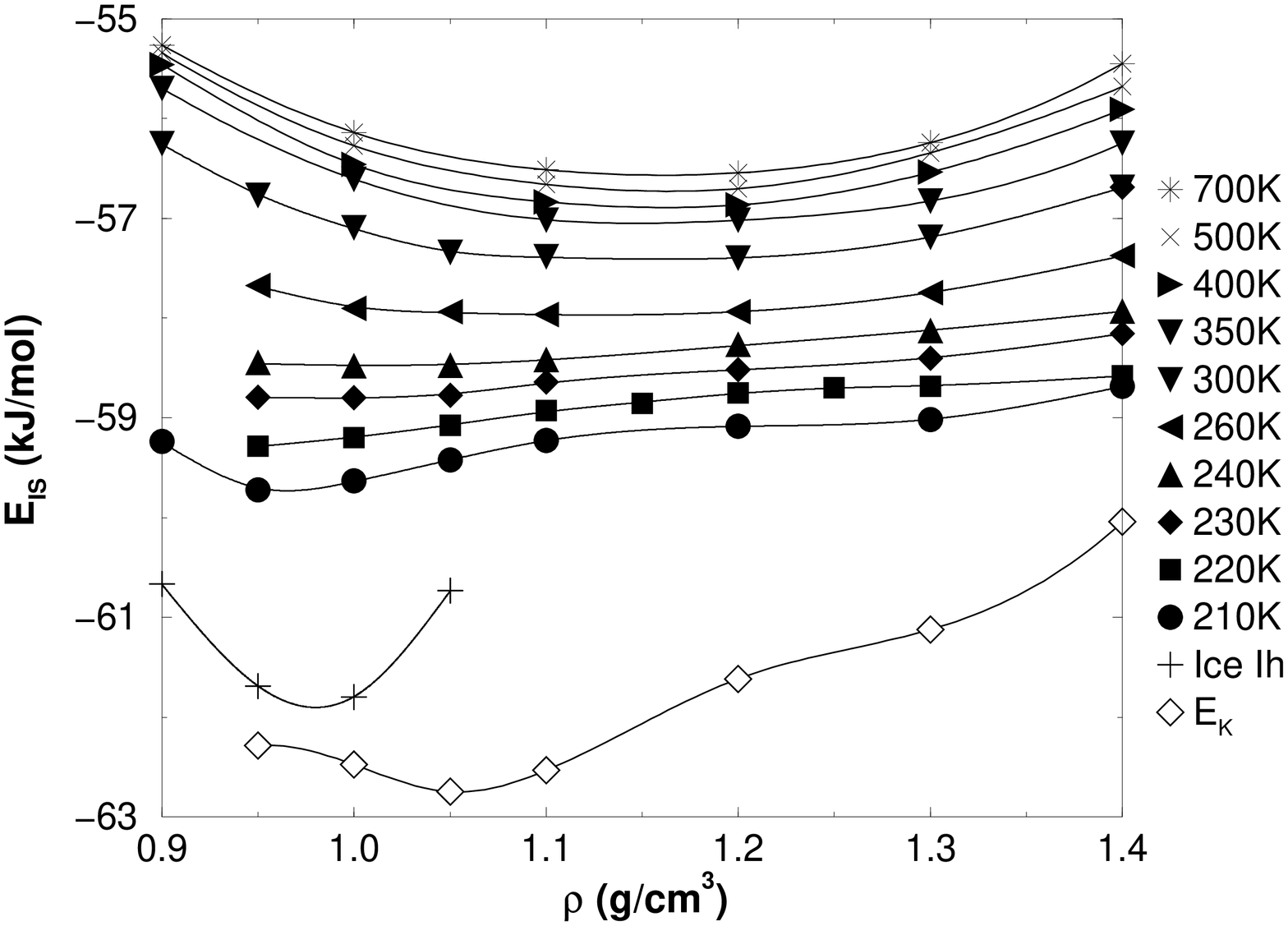,width=3.35in,angle=-90}}
\mbox{\psfig{figure=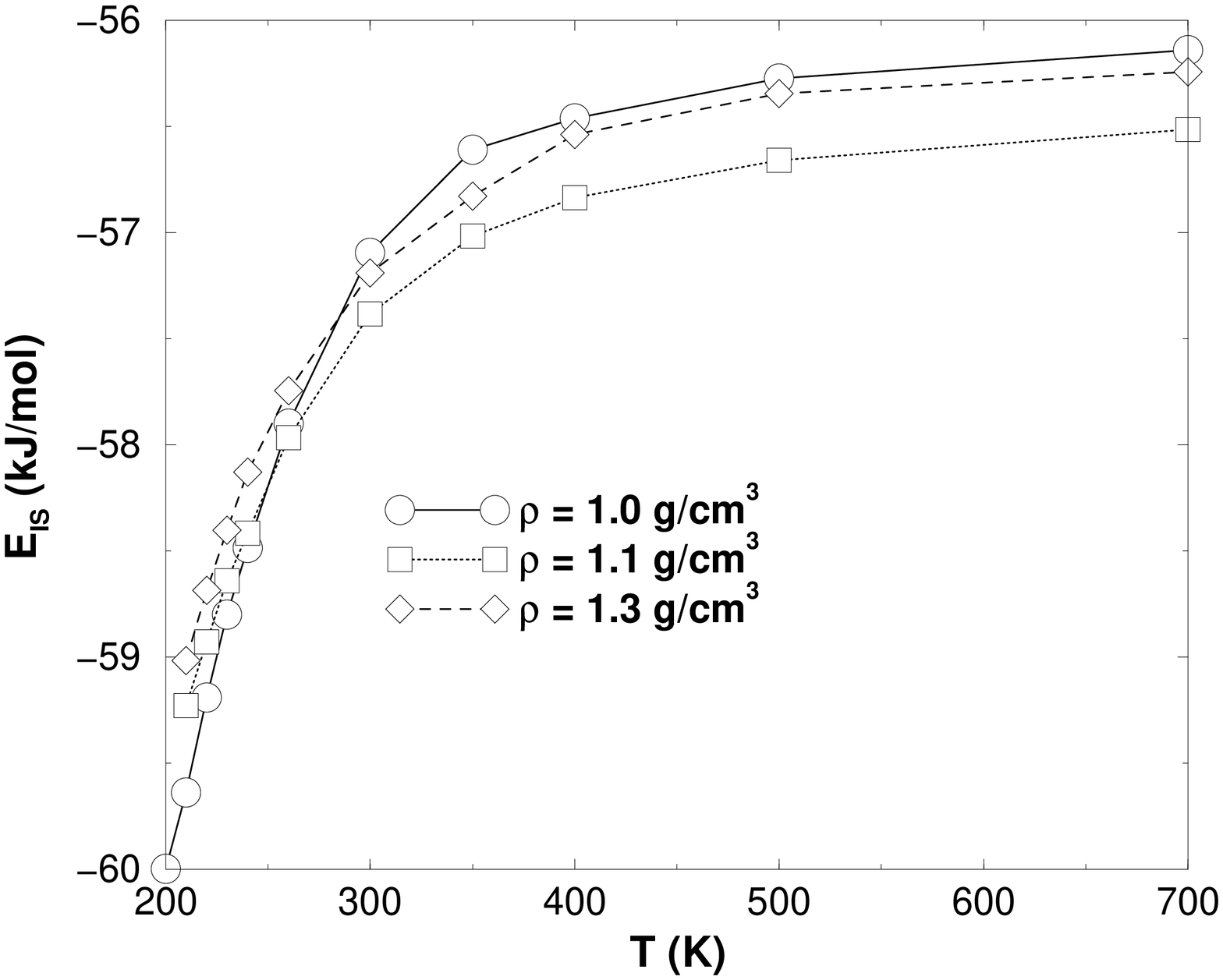,width=3.35in,angle=-90}}
\mbox{\psfig{figure=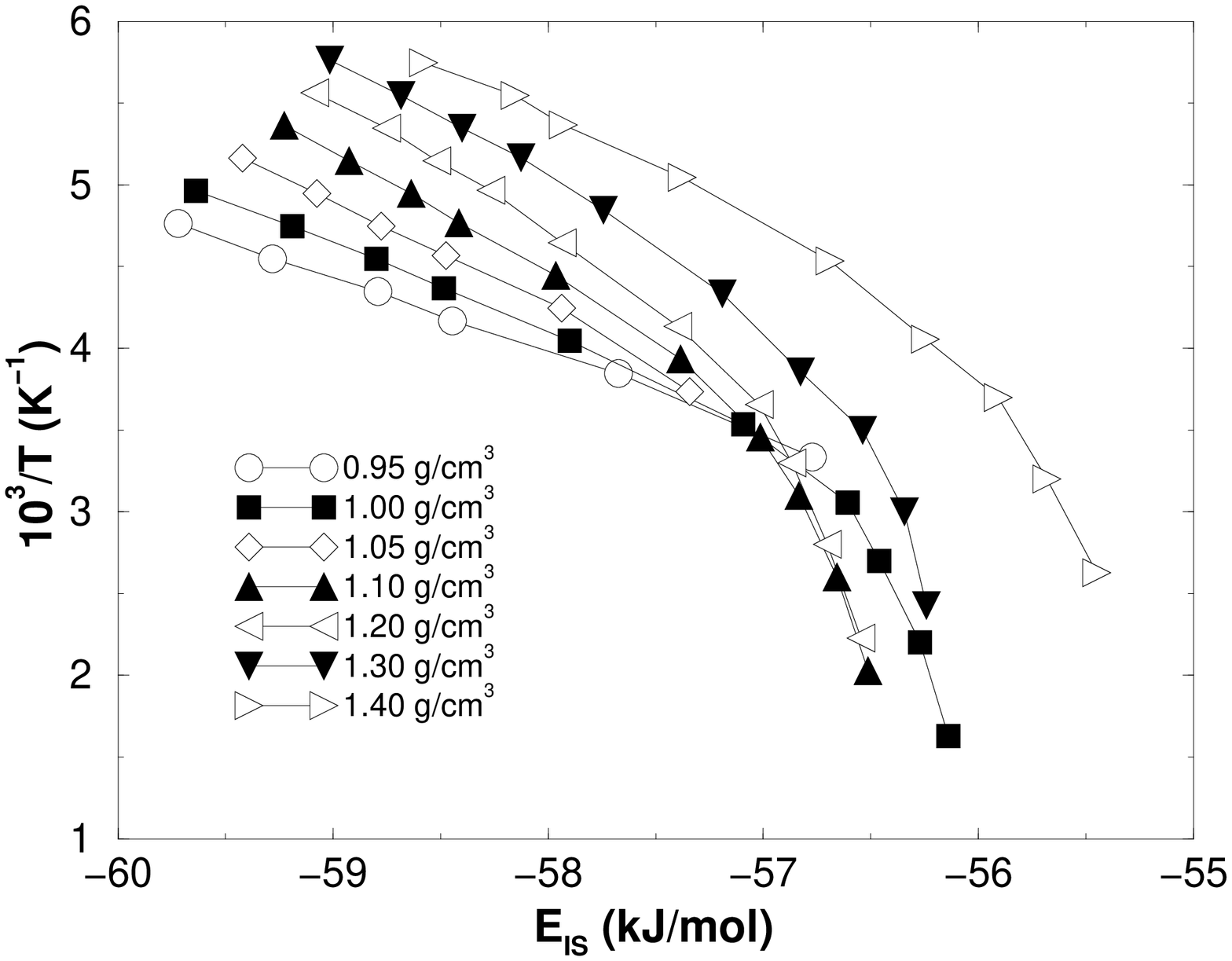,width=3.35in,angle=-90}}
\end{center}
\begin{minipage}{3.35in}
\caption{Inherent structure energy. (a) Density dependence of the
$E_{IS}$ for liquid configurations, and ice Ih. The bottom curve is
$E_K(\rho)$, obtained by fitting $S_{\mbox{\scriptsize conf}}$ using
Eq.~(\protect\ref{e:11}). Note that the three lowest isotherms, as well
as $E_K(\rho)$ display inflections.  The solid lines are shown as a
guide to the eye.  (b) $T$ dependence of $E_{IS}$ for selected
densities. For $T\protect\lesssim$~400~K, $E_{IS}$ starts to decrease
rapidly, corresponding to the fact that the systems populate basins of
lower energy. (c) $1/T$ vs $E_{IS}$ for all studied densities.  The
linear relationship at low $T$ support the possibility that the $e_{IS}$
values are Gaussian distributed.  Curves for different densities have
been shifted by multiplies of $0.2$ for clarity. }
\end{minipage}
\label{fig:ISenergy}
\end{figure}

The $\rho$- and $T$-dependence of $E_{IS}$ for the SPC/E potential for a
more limited range of $T$ was recently reported in Ref.~\cite{rds99}.
Our results for the IS energies are shown in Fig.~\ref{fig:ISenergy}(a)
as a function of $\rho$ and in Fig.~\ref{fig:ISenergy}(b) as a function
of $T$.

The IS energy of the liquid along isotherms shows hints of negative
curvature for $T \lesssim 230$~K (Fig.~\ref{fig:ISenergy}(a)).  The
presence of this negative curvature can also be observed in the
instantaneous equilibrium configurations \cite{hpss}. This curvature
yields a {\it negative\/} contribution to compressibility \cite{hpss}
$K_T^{-1} = V[(\partial^2U/\partial V^2]_T - T(\partial^2S/\partial
V^2)_T]$, which might be related to a low-temperature critical point in
SPC/E~\cite{spcet35}. Fig.~\ref{fig:ISenergy} also shows the IS energy
for ice $I_h$ (which coincides with the ground state energy).  At the
lowest $T$ studied, $E_{IS}$ of the disordered liquid is still
significantly greater than that of ice~\cite{badxtal}.  Note that ice
$I_h$ is not the thermodynamically stable crystalline form for SPC/E
\cite{baez}, and the stable SPC/E crystal would most likely have a lower
ground state energy.  However, the stable SPC/E crystal does not
correspond to any of the experimentally-known forms of
ice~\cite{ice-book}.  Fig~\ref{fig:ISenergy}(a) also show the Kauzmann
energy $E_K$, which we discuss in sec.~\ref{sec:entropy}.

Fig.~\ref{fig:ISenergy}(b) shows that at high $T$, $E_{IS}$ is nearly
$T$ independent.  For $T\lesssim 350-400$K, there is a rapid decrease of
$E_{IS}$ with a weak density-dependence.  At low $T$, $E_{IS}$ depends
linearly on $1/T$, as shown in Fig.~\ref{fig:ISenergy}(c). The $1/T$
dependence of $E_{IS}$ can be derived from the partition function
provided $\Omega(e_{IS})$ is Gaussian, and that the basin free energy
does not depend on $e_{IS}$~\cite{heuer,robin}.  Furthermore, as found
for the Lennard-Jones liquids~\cite{sri-nature}, the $T$ at which the IS
energy starts to decrease correlates with the temperature at which a
two-step relaxation starts to be observed in all characteristic
correlation functions (see, e.g., Fig.~13 of Ref.~\cite{shss})

\subsection{The Basin Free Energy: Density of States}

\begin{figure}[htbp]
\begin{center}
\mbox{\psfig{figure=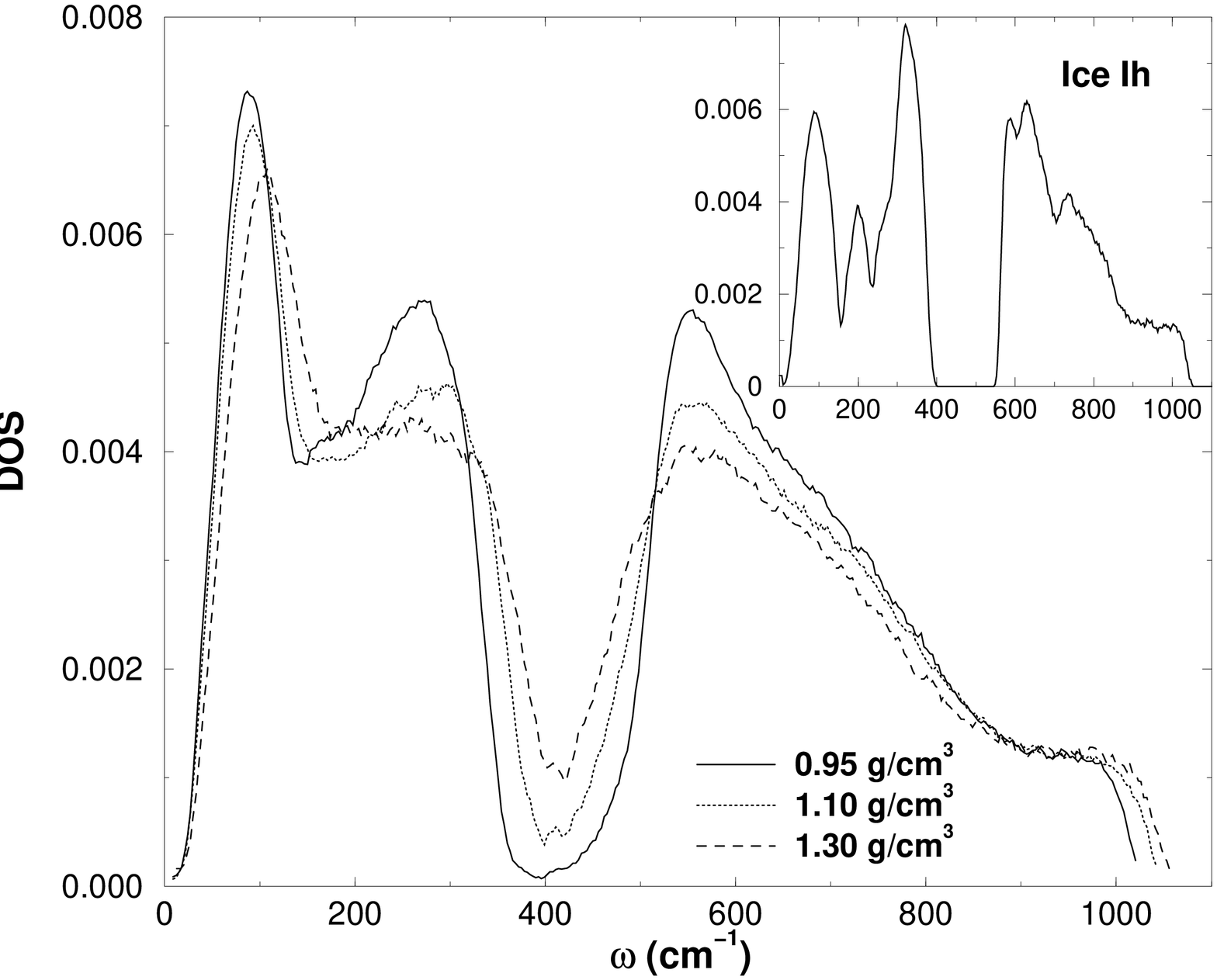,width=3.35in,angle=-90}}
\mbox{\psfig{figure=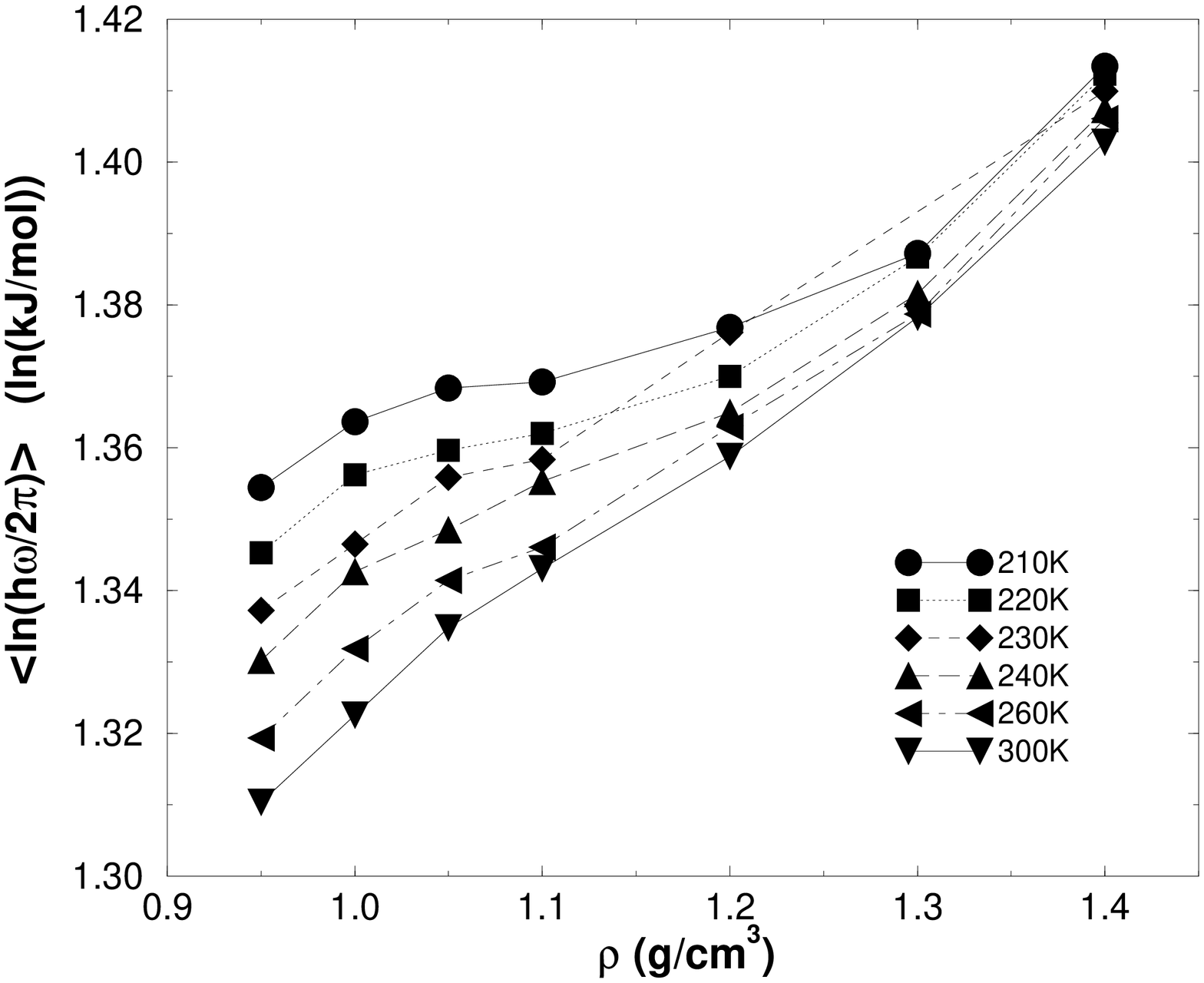,width=3.35in,angle=-90}}
\mbox{\psfig{figure=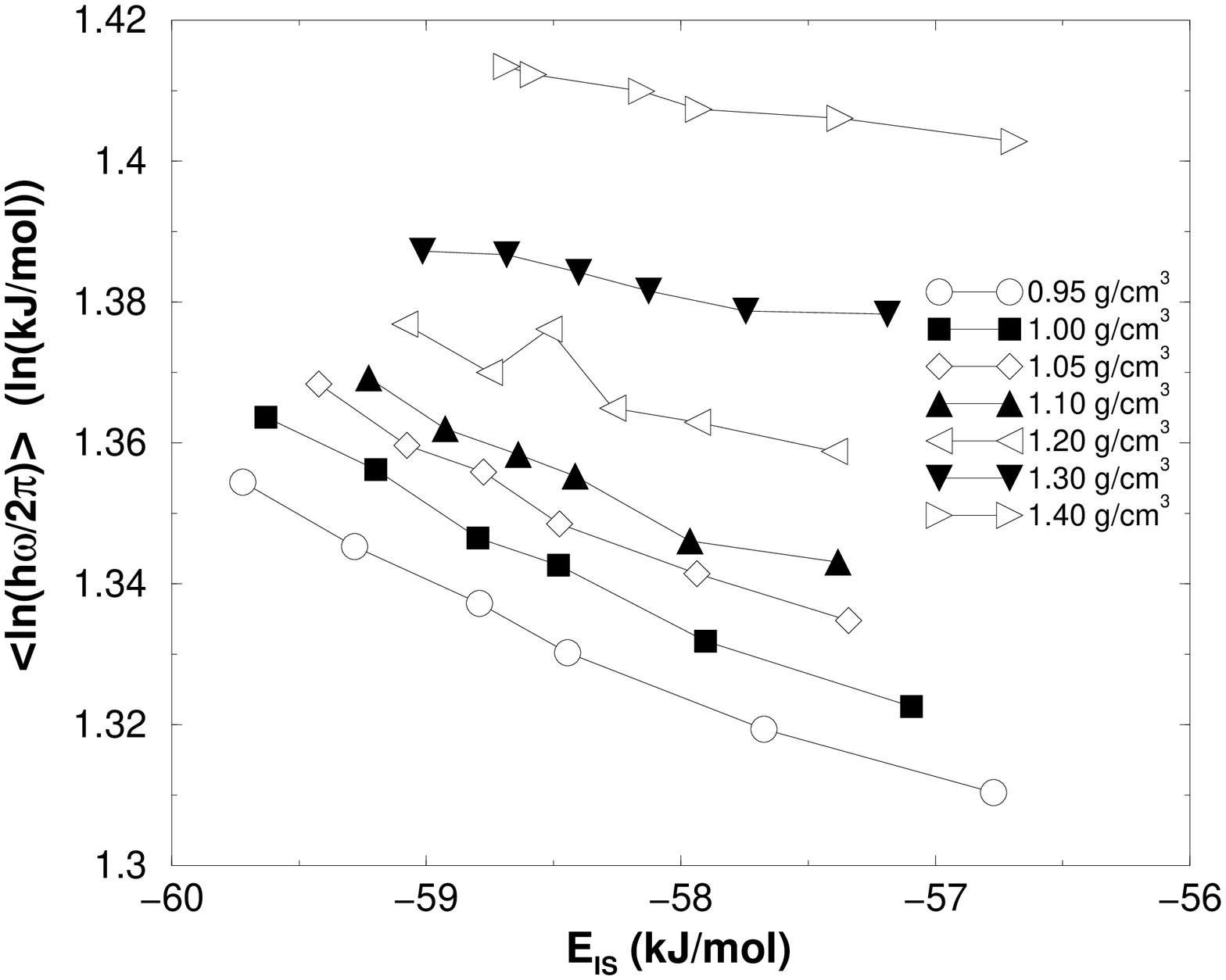,width=3.35in,angle=-90}}
\end{center}
\begin{minipage}{3.35in}
\caption{Shape of the basins surrounding the inherent structure. (a) The
DOS of the liquid at $T=210$~K.  The inset shows the DOS for ice Ih at
$\rho = 1.00$~g/cm$^3$ for comparison.  (b) Change of harmonicity of the
basins (c) $\langle\log(\hbar\omega)\rangle$ vs $E_{IS}$ for different
densities.}
\end{minipage}
\label{fig:basin-shape}
\end{figure}

We next focus on the shape of the IS basin with the aim of developing a
model for the basin free energy.  In the harmonic approximation, the basin
free energy is given by
\begin{equation} 
\label{e9}
F(E_{IS},T)=k_BT\sum_{i=1}^{6N-3}\left[\ln(\hbar\omega_i/k_BT)\right],
\end{equation}
the free energy of a harmonic oscillator with frequency spectrum
$\omega_i$. The values $\omega_i^2$ are the eigenvalues of the Hessian
matrix, defined by the second derivative of the potential energy with
respect to the molecular degrees of freedom at the basin minimum.  The
mass and moments of inertia of a molecule have also been absorbed in the
definition of $\omega_i$.  The distribution of $\omega_i$, called the
density of states (DOS), is shown in Fig.~\ref{fig:basin-shape}(a) for
three different state points at $T=210$~K.  The pronounced minimum at
$\omega \approx 400$~cm$^{-1}$ separates the translational modes (at
lower frequencies) from the rotational modes (at higher frequencies).
At larger $\rho$, the peaks in the DOS broaden due to the disruption of
the H-bond network, which we will discuss in Sec.~\ref{sec:structure}.
For comparison, we also show the DOS for ice $I_h$, where we see a clear
separation between the translational and rotational modes.

The normal mode spectrum contributes to the basin free energy via the
term $6\langle\ln(\hbar\omega)\rangle$ (per molecule), where the
brackets denote an average over the DOS and over different
configurations.  The dependencies on both $T$ and $\rho$ of
$\langle\ln(\hbar\omega)\rangle$ are shown in
Fig.~\ref{fig:basin-shape}(b). The dependence of $6k_B
\langle\ln(\hbar\omega)\rangle$ on $E_{IS}$ is shown in
Fig.~\ref{fig:basin-shape}(c). The average basin frequency is larger in
deeper basins, showing that the basins become increasingly ``sharp'' on
cooling.  This is in contrast with the Lennard Jones case, where the
basins become broader on cooling~\cite{eplaging,st-aging}.  The average
curvature of the IS basin at high density has a weaker
$E_{IS}$-dependence (and hence $T$-dependence) than those at low
density, but are generally larger than the curvature at low density.

\begin{figure}[htbp]
\begin{center}
\mbox{\psfig{figure=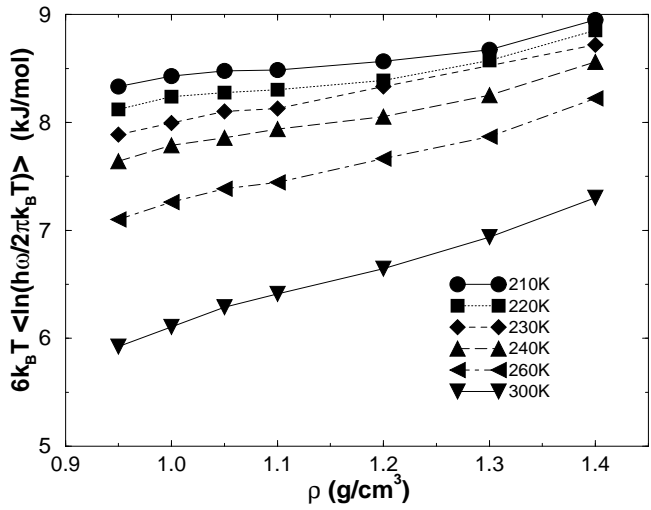,width=3.35in,angle=-90}}
\end{center}
\begin{minipage}{3.35in}
\caption{Free energy of a basin in the harmonic
approximation.}
\end{minipage}
\label{fig:hfeallT}
\end{figure}

Fig.~\ref{fig:hfeallT} shows the harmonic free energy estimate of
Eq.~(\ref{e9}) as a function of $\rho$. The range of values of the
vibrational free energy of Fig.~\ref{fig:hfeallT} (6 to 9 kJ/mol) is not
very different from the range of values of $E_{IS}$ of
Fig.~\ref{fig:ISenergy}(a) (from $-55$ to $-60$~kJ/mol); thus both make a
significant contribution to the free energy of Eq.~(\ref{e6}).

\begin{figure}[htbp]
\begin{center}
\mbox{\psfig{figure=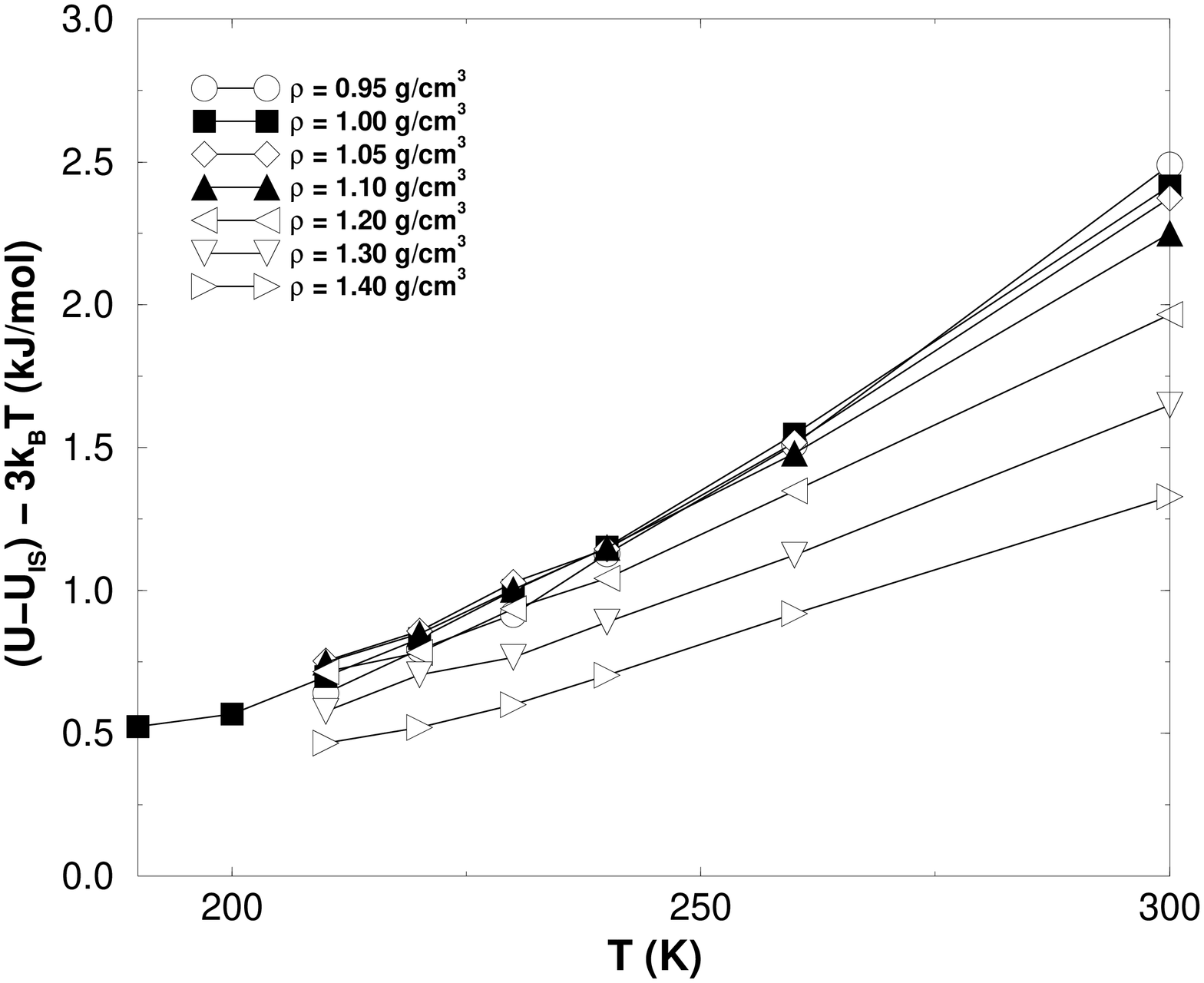,width=3.35in,angle=-90}}
\end{center}
\begin{minipage}{3.35in}
\caption{Anharmonicity of the basins as a function of $T$.}
\end{minipage}
\label{fig:anharmonicity}
\end{figure}

The basin free energy estimated using the harmonic approximation of
Eq.~(\ref{e9}) can be used as a starting point for estimating the true
basin free energy.  For a more precise quantification of the basin free
energy, we must consider anharmonic contributions to the free energy.
Indeed the basins of the SPC/E are anharmonic.  This can easily be seen
by considering the difference $U_{\mbox{\scriptsize vib}} \equiv
U-E_{IS}$; for a molecular system in the harmonic approximation,
$U_{\mbox{\scriptsize vib}} = (6/2)k_BT$.  Fig.~\ref{fig:anharmonicity}
shows a marked deviation from harmonic behavior.  However, as we discuss
next, the anharmonicity does not have a strong $E_{IS}$
dependence. Development of techniques for the estimation of the
anharmonic contribution to the basin free energy would be very useful.

\subsection{Basin Degeneracy}
\label{sec:entropy}
\begin{figure}[htbp]
\begin{center}
\mbox{\psfig{figure=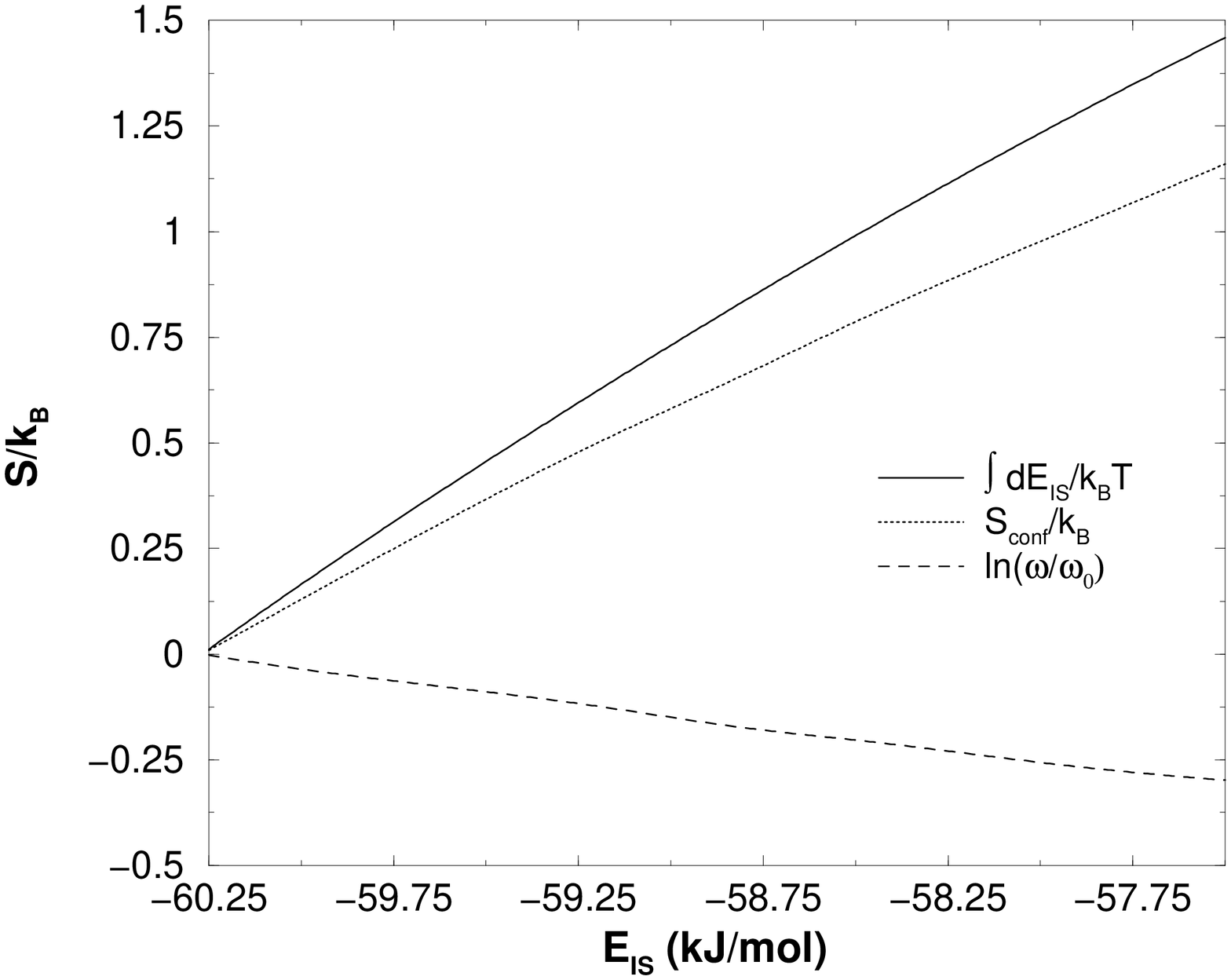,width=3.35in,angle=-90}}
\mbox{\psfig{figure=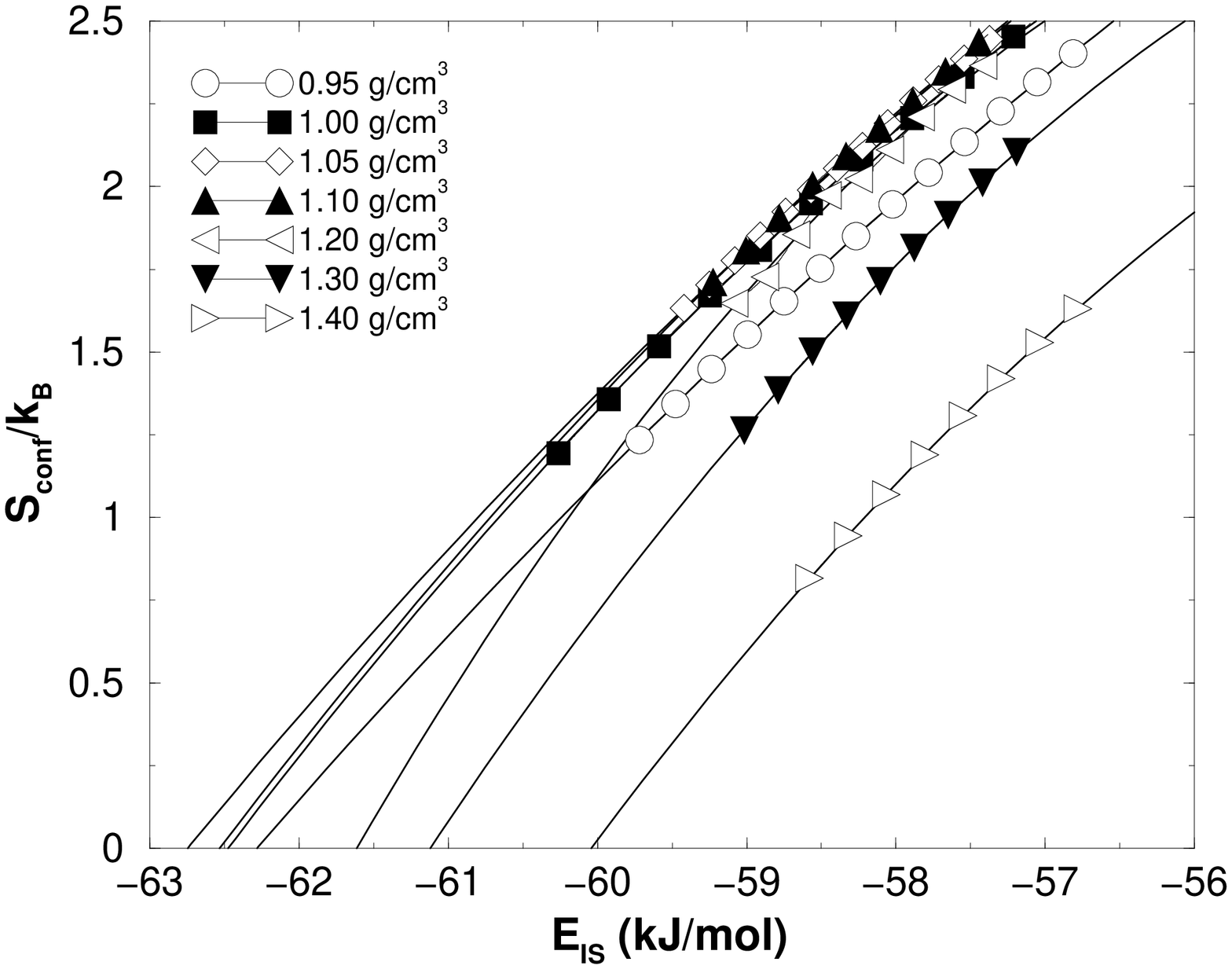,width=3.35in,angle=-90}}
\mbox{\psfig{figure=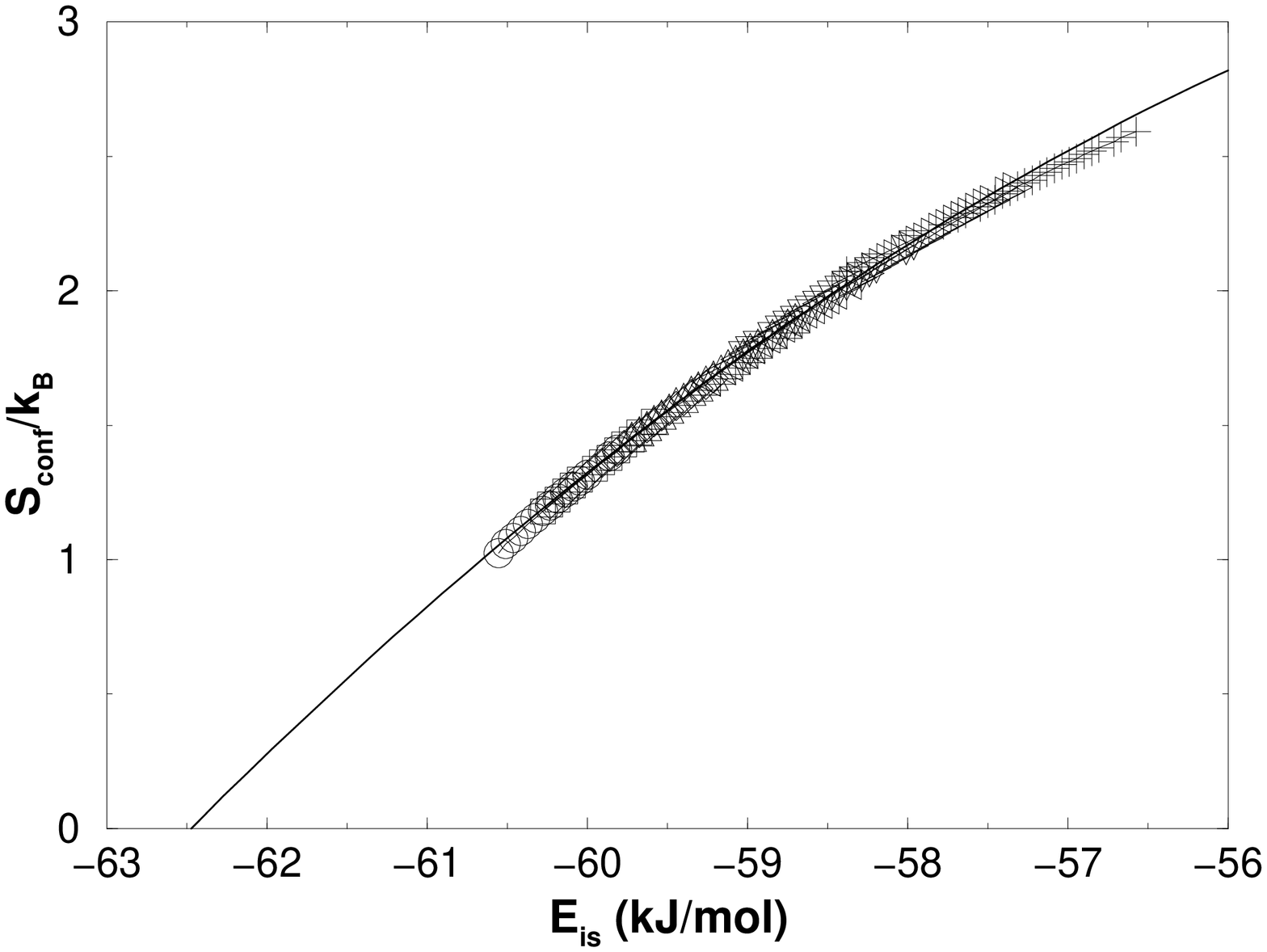,width=3.35in,angle=-90}}
\end{center}
\begin{minipage}{3.35in}
\caption{Configurational entropy $S_{\mbox {\scriptsize conf}}$. (a)
Role of the different contributions in the estimate of $S_{\mbox
{\scriptsize conf}}$.  (b) $E_{IS}$-dependence of the configurational
entropy for different $\rho$.  Lines are extrapolation to lower $E_{IS}$
value based on the assumption of linear relation between $E_{IS}$ and
$1/T$ at low $T$.  (c) Comparison between the $S_{\mbox{\scriptsize
conf}}$ values obtained using Eq.\protect\ref{eq:P(E_IS)} and
Eq.\protect\ref{eq:scalc} for $\rho=1.0$~g/cm$^3$.  Symbols refer to
different $P(E_{IS}(T),T)$ distributions, shifted to maximize the
overlap for different $T$-values.  }
\end{minipage}
\label{fig:hfevseis}
\end{figure}

A key element in the description of the configuration space in the IS
thermodynamic formalism is $\Omega(e_{IS})$, the number of basins with
energy $e_{IS}$. The corresponding configurational entropy $S_{\mbox
{\scriptsize conf}}$---i.e. the logarithm of $\Omega(e_{IS})$---can be
calculated by integrating Eq.~(\ref{eq:freeene}).  Using the harmonic
approximation of Eq.~(\ref{e9}) for the basin free energy, we obtain
\begin{eqnarray}
\label{eq:scalc}
S_{\mbox{\scriptsize conf}}(E_{IS})&=& S_{\mbox{\scriptsize conf}}(E_0) + 
\int_{E_0}^{E_{IS}} {{dE_{IS}}\over{T}}
+ \\ & & k_B \langle \ln(\omega(E_{IS})/\omega((E_{0})) \rangle .
\end{eqnarray}
Fig.~\ref{fig:hfevseis}(a) separately shows the contribution from $\int
dE_{IS}/T$, and the contribution associated with $\langle
\ln(\omega(E_{IS})/\omega((E_{0})) \rangle$. The harmonic contribution
is not negligible.  To obtain the $S_{\mbox {\scriptsize conf}}$ in
absolute scale, the value of $S_{\mbox {\scriptsize conf}}$ at the
reference point $E_0$ is needed.  We have used reference values obtained
in Ref.~\cite{sslss-nature} by independently calculating the absolute
value of the entropy via thermodynamic integration from the known ideal
gas reference point.

The complete results for all densities are shown in
Fig.~\ref{fig:hfevseis}(b). In the same graph we show a fit to
$S_{\mbox{\scriptsize conf}}(E_{IS})$ using the form
\begin{equation}
\label{e:11}
S_{\mbox{\scriptsize conf}}(E_{IS})= A (E_{IS}-E_{K})^2 + B (E_{IS}-E_{K})
\end{equation}
where $E_K$, the Kauzmann energy, is the $E_{IS}$ value at which the
configurational entropy vanishes.  This functional form is equivalent to
the Gaussian distribution of $\Omega(e_{IS})$~\cite{heuer}.  The
resulting $E_{IS}$ values, which provide an indication of the $\rho$
dependence of $E_K$, are reported in Fig.~\ref{fig:ISenergy}(a).  The
$E_K$ values suggest that disordered states with energies comparable to
the ordered crystalline states may be available to this
system~\cite{rds99}, provided the $E_K$ estimates are not significantly
affected by unknown errors in the extrapolation procedure or in the
reference value for $S_{\mbox{\scriptsize conf}}(E_0)$ used.

To confirm independently the validity of the approach followed to
estimate $E_{IS}$ dependence of $S_{\mbox{\scriptsize conf}}$, we
compare the value obtained from Eq.~(\ref{eq:scalc}) with the value
obtained using Eq.~(\ref{eq:P(E_IS)}) in Fig.~\ref{fig:hfevseis}(c). The
agreement between the curves is remarkable.  Moreover, the overlap for
different $P(e_{IS},T)$ distributions after the harmonic $E_{IS}$
dependence is taken into account, suggests that there are no other
systematic $E_{IS}$-dependent contributions. Any remaining $T$-dependent
contributions are absorbed in the unknown $Z(T)$ function.

The results reported in this section provide a detailed analysis of the
inherent structures and of their basins. This information can be used to
develop a detailed free energy expression for the SPC/E~\cite{spcet35}.
Calculation of the $T$ dependence of $S_{\mbox {\scriptsize conf}}$ was
presented in Ref.~\cite{sslss-nature}, to probe the relation between the
configurational contributions to thermodynamic quantities and the liquid
dynamics in supercooled states.

\section{Structure} 
\label{sec:structure}

On lowering $T$ the dramatic changes in the IS energies are known to be
accompanied by equally dramatic changes in the dynamic properties of the
instantaneous configurations.  However, examination of simple measures
of structure of the instantaneous configurations (such as the pair
distribution function) do not reveal such obvious changes; rather, there
is a very gradual change, with the structure becoming slowly more well
defined.  A more careful analysis of structure is is required to see
significant changes~\cite{jonsson}.  In the following, we focus our
attention on the structural changes that can be observed by studying the
IS to try to obtain a more clear picture of the structural evolution of
the system on cooling.  The results we present are complementary to the
results recently reported for the same system along similar lines of
thought~\cite{rds99} and expand on previous work
\cite{sciortino2,starrproc}.

\subsection{Pair distribution function}
\begin{figure}[htbp]
\begin{center}
\mbox{\psfig{figure=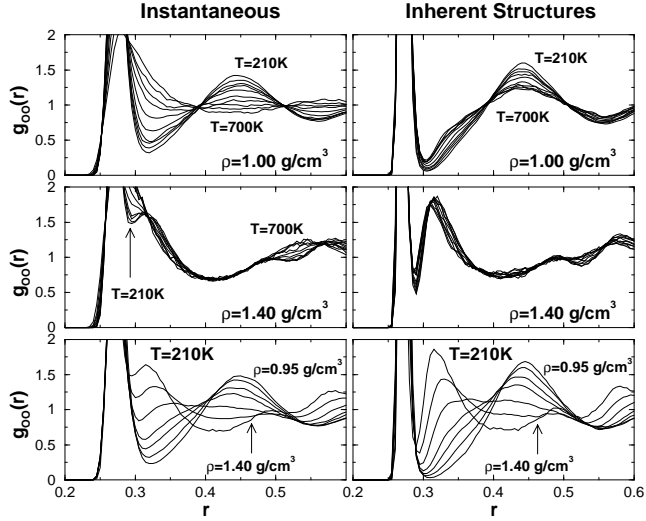,width=3.35in,angle=-90}}
\end{center}
\begin{minipage}{3.35in}
\caption{Oxygen-oxygen pair correlation function, for equilibrated
liquid configurations, and inherent structures. Upon lowering
temperature, the first peak in both cases becomes sharper, and the
intensity between the first and second peaks decreases. The smaller
changes seen in the case of the inherent structures offers an estimate
of that part of the change due to configurational change upon cooling, as
opposed to thermal effects.}
\end{minipage}
\label{fig:gr}
\end{figure}

\begin{figure}[htbp]
\begin{center}
\mbox{\psfig{figure=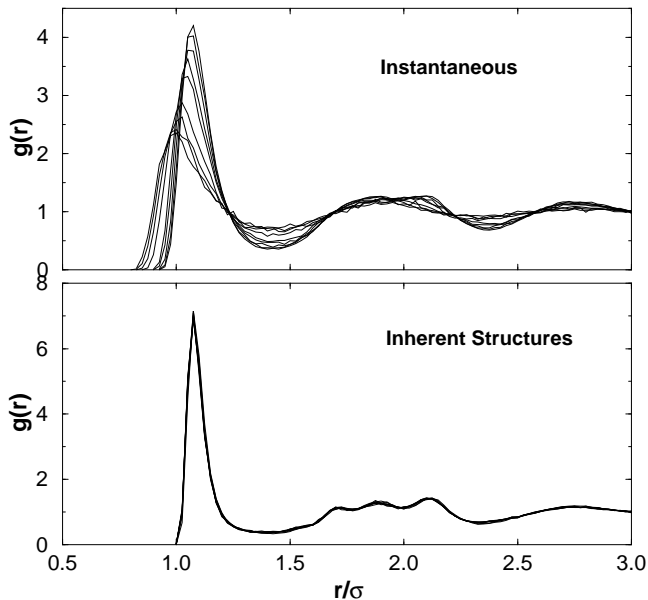,width=3.35in,angle=-90}}
\end{center}
\caption{Pair correlation function for equilibrated liquid
configurations, and inherent structures of the frequently studied binary
Lennard-Jones mixture between $T=0.44$ and $T=5$ in units of the pair
potential energy depth [the model parameters are defined in
Ref.~\protect\cite{skt}].  Upon lowering temperature, there is no
apparent change in the structure of the IS, similar to the behavior we
observe for water at large $\rho$.}
\label{fig:ljgr}
\end{figure}

Fig.~\ref{fig:gr} shows the oxygen-oxygen pair correlation function for
both the equilibrated liquid and for the inherent structures at various
$T$ and $\rho$.  We first focus our attention on the $T$ dependence
along the $\rho = 1.00$~g/cm$^3$ isochore.  For equilibrated
configurations as well as inherent structures, the first and second
peaks of the pair correlation function become better defined upon
decreasing the temperature. Further, there is a systematic reduction of
the intensity between the first and second neighbor peaks.  At higher
density ($\rho = 1.40$~g/cm$^3$), the behavior is somewhat different;
for the instantaneous configurations, there is a clear $T$ dependence,
while for the IS, the $T$ dependence is nearly negligible.  For
comparison, the model atomic liquid studied in~\cite{sri-nature,skt},
shows virtually no $T$ dependence of $g(r)$ in the IS
(Fig~\ref{fig:ljgr}).  Hence the behavior at large $\rho$ is more like
that expected for a simple liquid, consistent with the disappearance at
large $\rho$ of many of the anomalies that differentiate water at
ambient density.

We also show the behavior of $g(r)$ for various $\rho$ at $T=210$~K
in the bottom panels of Fig.~\ref{fig:gr}.  On increasing $\rho$, both
the instantaneous and inherent structure configurations show an increase
in intensity at $r \approx 0.32$~nm, a trait already known to develop
due to distortion (and eventual interpenetration) of the hydrogen bond
network.  For water, it is known that the preferred nearest-neighbor
geometry is tetrahedral.  This tetrahedral ordering is obvious at low
densities, where peaks at $0.28$~nm and $0.45$~nm are the expected
distances for a perfect tetrahedral lattice with first neighbor
separation of $0.28$~nm.  To see more clearly the tetrahedral nature of
the liquid at all these temperatures and densities, we focus on the
neighbor statistics at the various state points.

\subsection{Neighbor Changes}

We first consider the the average number of neighbors that a molecule
has within a sphere of radius $r$, which is obtained from integration of
$g(r)$:

\begin{equation}
n(r) = 4\pi\rho \int_0^r r'^2 g(r') dr'.
\end{equation}

\begin{figure}[htbp]
\begin{center}
\mbox{\psfig{figure=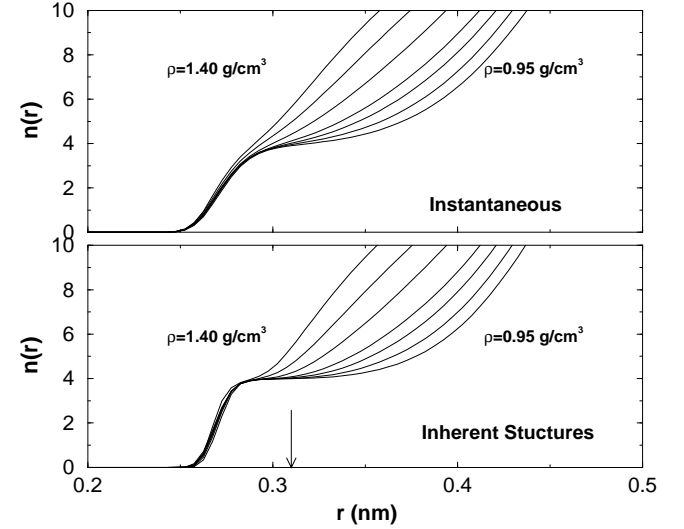,width=3.35in,angle=-90}}
\end{center}
\caption{Average number of neighbors $n(r)$ as a function of distance
$r$.  For all densities, an average of nearly four neighbors is reached
at the first neighbor distance, showing the tendency for short range
tetrahedral ordering, even at very high density.}
\label{fig:n(r)}
\end{figure}

\noindent We show $n(r)$ in Fig.~\ref{fig:n(r)} as a function of $\rho$
at the lowest $T$ studied.  We see a plateau in $n(r)$ almost exactly
equal to four for all densities.  Therefore, even at large $\rho$, the
liquid has short range tetrahedral order.  However, the rapid growth of
$n(r)$ at large $\rho$ highlights the distortion and interpenetration.

\begin{figure}[htbp]
\begin{center}
\mbox{\psfig{figure=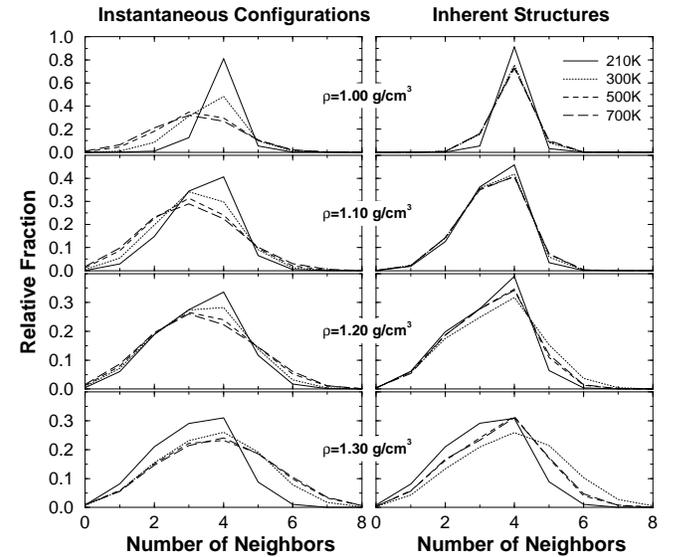,width=3.35in,angle=-90}}
\end{center}
\caption{\narrowtext Distribution of the coordination number of
molecules for both the instantaneous equilibrated configurations, and
the inherent structures. In all cases, the fraction of four-coordinated
molecules increases with decreasing $T$.}
\label{fig:bond-dist}
\end{figure}

To quantify the $T$ dependence of the structural changes we calculate
the distribution of the number of neighbors a molecule has within a
distance $0.31$~nm (arrow in Fig.~\ref{fig:n(r)}), roughly corresponding
to the first minimum in $g(r)$ at low density, shown in
Fig.~\ref{fig:gr}.  We choose the first minimum in $g(r)$ at low density
to emphasize the tetrahedrality of the liquid; for ice, all molecules
would have 4 neighbors, while in high density liquid configurations, the
distortion leads to a significant number of molecules having more or
less than 4 neighbors in the first shell.  Fig.~\ref{fig:bond-dist}
shows the histograms of the fraction of molecules with a given
coordination number. At low density, the histogram for the instantaneous
configurations changes from a rather broad one to one that is peaked
around the value $4$ as the temperature decreases, as we expect.  The
same trend is visible for the inherent structures, although even at high
temperatures, the distribution is quite narrowly peaked around the value
$4$.  Such a comparison permits us to make a separation between
deviations from four-coordination arising from thermal agitation, and
that arising from configurational change.  This less marked $T$
dependence of the IS can be seen for all $\rho$.  At larger $\rho$, the
distribution is still peaked around 4, but there is significant fraction
of molecules with are not four-bonded --- these are molecules are most
likely participating in the so called {\it bifurcated
bonds}~\cite{sciortino2}.

\begin{figure}[htbp]
\begin{center}
\mbox{\psfig{figure=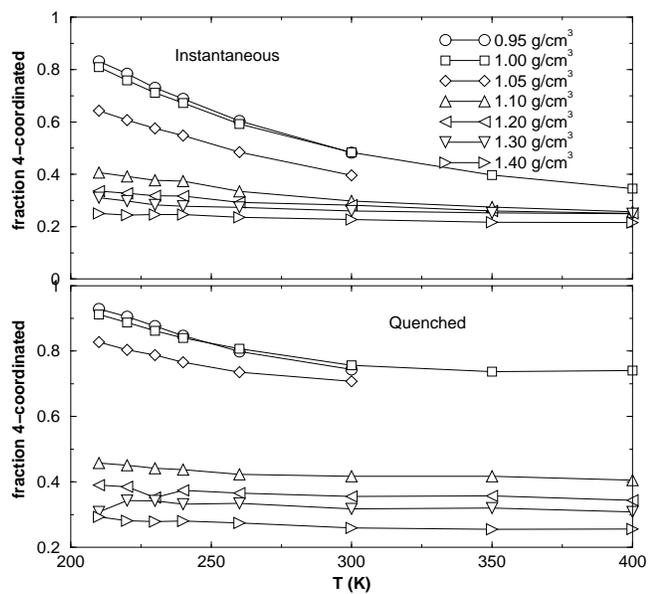,width=3.35in,angle=-90}}
\end{center}
\caption{Fraction of four coordinated water molecules: (a) equilibrated
configurations and (b) inherent structures.}
\label{fig:4bonds}
\end{figure}

\begin{figure}
\begin{center}
\mbox{\psfig{figure=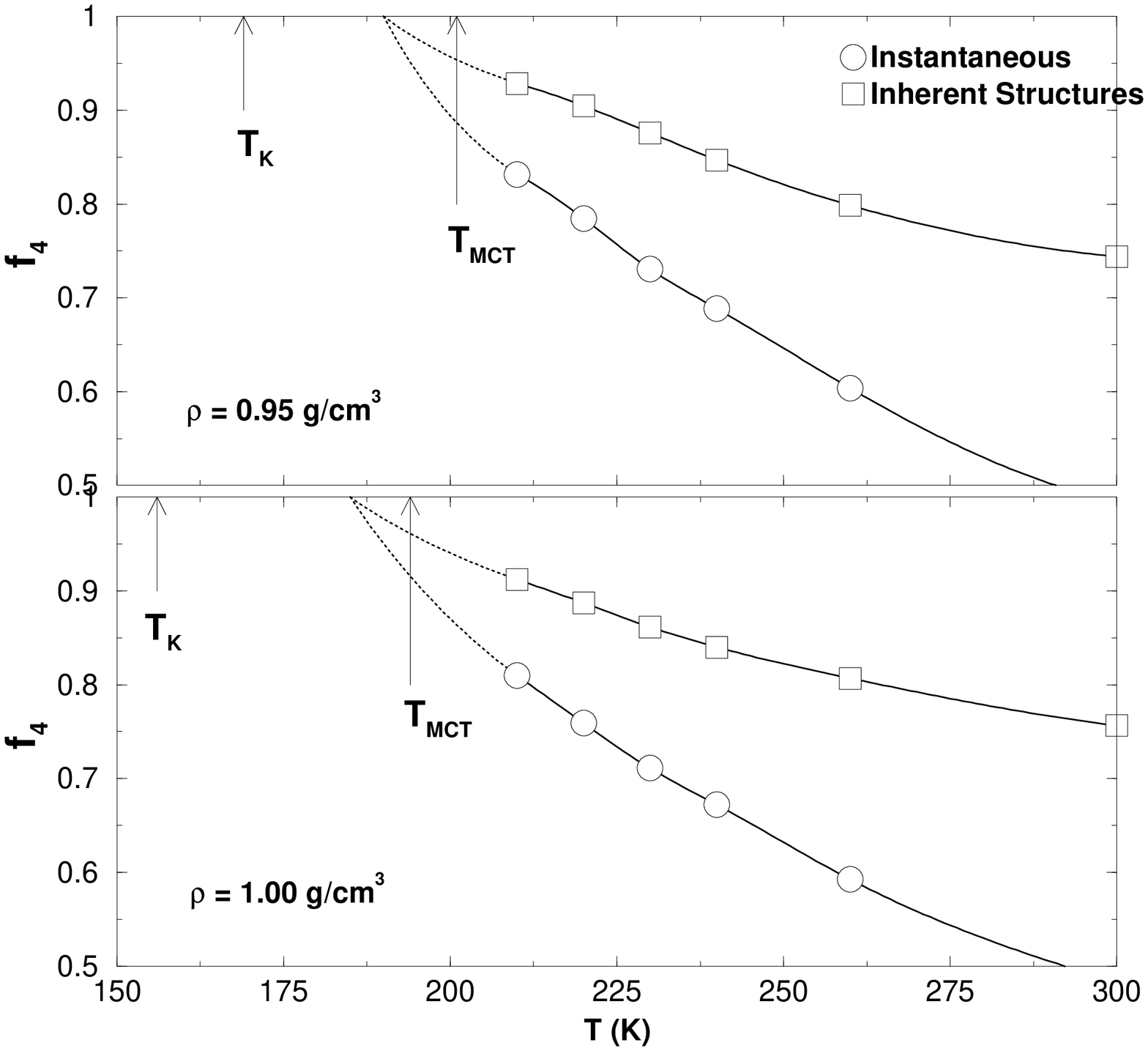,width=3.35in,angle=-90}}
\end{center}
\caption{Extrapolation for $f_4$ to lower $T$ to estimate $T(f_4)=1$.}
\label{fig:f4=1}
\end{figure}

To quantify the changes in the tetrahedrality as a function of $T$ for
each $\rho$, we plot the fraction of four-bonded molecules $f_4$ as a
function of $T$ in Fig.~\ref{fig:4bonds}.  Those molecules which are not
four bonded represent the set of bifurcated bonds~\cite{sciortino2}.  At
low density, it is interesting to note that for both the equilibrated
configurations and inherent structures, this $f_4$ is close to $1$ at
the lowest $T$ simulated.  Indeed, for $T$ somewhat lower that we can
currently simulate, it appears that all molecules would be four bonded,
and hence form a perfect random tetrahedral network.  A simple
extrapolation of $f_4$ for the lowest densities, displayed in
Fig.~\ref{fig:f4=1}, shows that $T(f_4=1)$ appears slightly lower than
the mode-coupling transition temperature $T_{MCT}$
\cite{shss,workonMCT,fs-review} but well above $T_K$.  Additionally, it
appears that both the instantaneous and IS configurations appear to
reach a random tetrahedral network at roughly the same temperature.  The
close correspondence of $T(f_4=1)$ and $T_{MCT}$ suggests that this
crossover in structural change may be the controlling mechanism for the
crossover in dynamic properties at $T_{MCT}$~\cite{tbs,st,lanave}.
However, the extrapolation does not allow us to unambiguously associate
these temperatures.  Since the tetrahedral geometry is the ``ideal''
configuration at these low densities, there are unlikely to be any
significant structural changes in the liquid for $T<T(f_4=1)$; to the
extent that there is further structural change for $T<T(f_4=1)$, the
rate of change must be significantly different than for
$T>T(f_4=1)$. Analysis of the neighbor statistics in the IS at lower $T$
might shed some light on the hypothesized change in the dynamics from
that of a fragile liquid to a strong liquid\cite{itoet,starr,swedes}, a
widely debated topic~\cite{bruce}.

At larger densities increases, it is apparent that a random tetrahedral
network is never reached.  This is apparent in the fact that there are
far fewer four-bonded molecules in the first neighbor shell at the
higher densities.  Hence, no sharp change in the structural development
as a function of $T$ is expected.  Consequently any crossover from
fragile to strong behavior must become less pronounced.
\section{Conclusions}

We have characterized the properties of the PES basins in the
configuration space of the SPC/E model over a range of densities in
the supercooled regime.  We have shown that the $T$ dependence of
$E_{IS}$ is qualitatively very similar to the previous observations of
simple liquids.  We have also shown the importance of accounting for
the shape of the basins when characterizing the thermodynamic
properties of the IS subsystem.  In particular, we found the there are
significant $T$ dependent anharmonic effects.  This detailed
information on the basins should be useful for any future studies that
focus on the dynamics of exploring the PES, or the thermodynamics at
low $T$.  Such a classification of basin properties would also be
useful for other model glass formers, and might help to highlight the
differences between fragile (such as OTP) and strong (such as SiO$_2$)
glass formers\cite{fragility}.

In addition, we presented results for the structural changes of the
IS; these changes are more pronounced than what can be seen from
equilibrium configurations.  In particular, the progression of the
structure toward a random tetrahedral network at low $T$ holds promise
for a more physical and intuitive understanding of the glass transition
in water.  Below the temperature of crossover to a tetrahedral network,
no further structural arrangement is expected, and the configurational
entropy of the liquid may be nearly ``frozen in'' at the value that
corresponds to the random tetrahedral network (plus the residual defects
that would be present at concentrations varying with density). Thus the
rate of change in entropy of the liquid may be expected change
substantially near the crossover temperature, resulting in a
significantly lower value of $T_K$, compared to the value that may be
expected if the rate of change above the crossover temperature
persisted. Similarly, because of the significant temperature dependence
of the fraction of bifurcated bonds -- which facilitate structural
rearrangement -- above the cross-over temperature, and the relative
constancy below, the temperature dependence of the dynamical properties
may show a corresponding crossover.

Finally, we call attention on the possibility of studying physical aging
in this model-system, starting from the thermodynamic description of
this system.  We note that, at $\rho = 1.40$~g/cm$^3$, there is almost
no variation of the basin curvature (Fig.~\ref{fig:basin-shape}) on
$E_{IS}$, nor does the structure change significantly
(Fig.~\ref{fig:gr}).  Hence these high density state points may offer an
ideal opportunity to check if the out-of-equilibrium dynamics at very
low $T$ can be still related to equilibrium states of the
system~\cite{st-aging}.

\section{Acknowledgments} We thank P.G.~Debenedetti, S.G.~Glotzer,
and F.H.~Stillinger for helpful discussions.  F.W.S. is supported by the
National Research Council.  F.S. is supported by INFM-PRA-HOP and {\it
Iniziativa Calcolo Parallelo} and from MURST PRIN 98.  This work was
also supported by the NSF.

\begin{minipage}{3.35in}
\begin{table}
\caption{Summary of the thermodynamics properties obtained from
simulations of Ice Ih, all at $T = 194$~K.}
\medskip
\begin{tabular}{c|c|c}
$\rho$ (g/cm$^3$) & P (MPa) & U (kJ/mol) \\
\tableline
0.90 & -560 & -55.52 \\
0.95 & -77  & -56.38 \\
1.00 & 475  & -56.34 \\
1.05 & 700  & -55.44
\end{tabular}
\label{table:ice-simulations}
\end{table}
\end{minipage}

\end{multicols}


\begin{references}
\bibitem{badxtal} In ref.~\cite{rds99} the IS energy of the liquid at
230~K was found to be lower than that of the estimated ice Ih ground
state energy for $\rho \lesssim 0.95$~g/cm$^3$ and $\rho \gtrsim
1.05$~g/cm$^3$.  In contrast, our results show that ice Ih is at a
significantly lower energy than than the liquid IS at a lower
temperature of 210~K.  This is most likely explained by the fact that
our results are direct calculation of the IS energy, while the results
of Ref.~\cite{rds99} are only an estimation of the ground state
energies.
\end{references}
\end{document}